\documentclass{article}

\usepackage{arxiv}

\usepackage[utf8]{inputenc} 
\usepackage[T1]{fontenc}    
\usepackage{hyperref}       
\usepackage{url}            
\usepackage{booktabs}       
\usepackage{amsfonts}       
\usepackage{nicefrac}       
\usepackage{microtype}      
\usepackage{lipsum}		
\usepackage{graphicx}
\usepackage[square,sort,comma,numbers]{natbib}
\usepackage{doi}
\usepackage{amssymb}
\usepackage{latexsym}

\usepackage{subcaption}
\usepackage{makecell}

\title{Pristine annotations-based multi-modal trained artificial intelligence solution to triage chest X-Ray for COVID19}

\author{ \href{}{\hspace{1mm}Tao Tan \thanks{tao.tan@ge.com} \qquad Bipul Das  \qquad Ravi Soni  \qquad Mate Fejes} \\
  \\
\And
  Sohan Ranjan  \qquad Daniel Attila  Szabo \qquad Vikram Melapudi \qquad K S Shriram \\
  \\
\AND
Utkarsh  Agrawal \qquad Laszlo Rusko \qquad Zita Herczeg \qquad Barbara Darazs \\
 \\
 \AND
Pal Tegzes \qquad Lehel Ferenczi\qquad Rakesh Mullick \qquad Gopal Avinash\\ 
GE healthcare \\
}

\renewcommand{\shorttitle}{\textit{arXiv} Template}

\hypersetup{
pdftitle={A template for the arxiv style},
pdfsubject={q-bio.NC, q-bio.QM},
pdfauthor={David S.~Hippocampus, Elias D.~Striatum},
pdfkeywords={Multi-modal, COVID-19, Artificial intelligence},
}

\begin{document}

\graphicspath{ {./figure/} }

\maketitle

\begin{abstract}
The COVID-19 pandemic continues to spread and impact the well-being of the global population. The front-line modalities including computed tomography (CT) and X-ray play an important role for triaging COVID patients. Considering the limited access of resources (both hardware and trained personnel) and decontamination considerations, CT may not be ideal for triaging suspected subjects. Artificial intelligence (AI) assisted X-ray based applications for triaging and monitoring which require experienced radiologists to identify COVID patients in a timely manner and to further delineate the disease region boundary are seen as a promising solution. Our proposed solution differs from existing solutions by industry and academic communities, and demonstrates a functional AI model to triage by inferencing using a single x-ray image, while the deep-learning model is trained using both X-ray and CT data.  We report on how such a multi-modal training improves the solution compared to X-ray only training. The multi-modal solution increases the AUC (area under the receiver operating characteristic curve) from 0.89 to 0.93 and also positively impacts the Dice coefficient (0.59 to 0.62) for localizing the pathology. To the best our knowledge, it is the first X-ray solution by leveraging multi-modal information for the development.	
\end{abstract}


\section{Introduction}

Coronavirus disease 2019 (COVID-19) is extremely contagious and has become a pandemic \cite{al2020death,shaker2020covid}. It has spread inter-continentally in the first wave \cite{leung2020first} and suspected to be currently entering the second wave \cite{ali2020covid, aleta2020modeling, evenett2020preparing} in various countries, having infected more than 30 million people and caused nearly 1M deaths till September 2020 \cite{JHU}. The mortality rate of this disease differs from country to country ranging from 2.5\% to 7\% compared with 1\% from influenza \cite{baud2020real, rajgor2020many, jung2020real}. Considering different age groups, elder people and patients with comorbidities are most vulnerable and more likely to progress to a life-threatening condition\cite{liu2020clinical}. To prevent the spread of the disease, different governments have implemented strict containment measures \cite{tian2020investigation,gatto2020spread} which have aimed to minimize transmission. Because of the strong infection rate of COVID-19, rapid and accurate diagnostic methods are urgently required to identify, isolate and treat the patients specially considering that effective vaccines are still under development.

The diagnosis of COVID-19 relies on reverse-transcriptase-polymerase chain reaction (PCR) test \cite{lanciotti1992rapid, kim2020diagnostic, kucirka2020variation}. However it has several drawbacks. The PCR tests often require 5 to 6 hours to yield results. Sensitivity of PCR depends on the stage of the infection \cite{kucirka2020variation} and can be as low as 71\%\cite{fang2020sensitivity}. Therefore care must be taken on interpreting PCR tests. More importantly, the cost of PCR prevents large population from being tested in the developing and highly populated economies.

From the imaging domain, chest CT may be considered as a primary tool for the current COVID-19 detection \cite{ai2020correlation, kim2020diagnostic, udugama2020diagnosing} and the sensitivity of chest CT can be greater than that of PCR (98\% vs 71\%) \cite{fang2020sensitivity}, but the cost and time including for system decontamination can be prohibitive. Portable X-ray (XR) units are cost and time effective, and thus is the primary imaging modality used in diagnosis and management. Since it has limited capability to provide detailed 3D structure of anatomy or pathology of chest cavity and therefore not regarded as an optimum tool for quantitative analysis \cite{blanchon2007baseline}. Also due to the imaging apparatus and nature of the x-ray projection, it is challenging to the radiologists to identify relevant disease regions and creates difficulty in accurate interpretation \cite{jacobi2020portable, pereira2020covid}.

To alleviate the lack of experienced radiologists and minimize human effort in managing an exponentially growing pandemic  and the impending task to triage suspected COVID-19 subjects, the academic and industry communities have proposed various systems for diagnosing COVID-19 patients using X-ray imaging \cite{castiglioni2020artificial, murphy2020covid, shi2020review, neri2020use}. The performance of some AI systems to detect COVID-19 pneumonia was comparable to radiologists to identify presence or absence of COVID-19 infection\cite{murphy2020covid}. The common outputs of these diagnostic solutions or triage systems are classification likelihoods for COVID19 and non-COVID-19 accompanied by heatmaps localizing the suspected pathological region or areas of attention.

Although the performance of some systems approaches the level of radiologists on X-rays in terms of classification, but to our best knowledge, no studies have verified the detection and segmentation of the disease regions against human annotation on X-rays. Segmentation is critical for (a) severity assessment of the disease; and for (b) follow-up for treatment monitoring or progression of patient condition. Although human annotations can be obtained for COVID-19 regions on X-rays, the certainty of regions is weak compared to annotations derived from CT. The development in the field suffers from the lack of availability of COVID-19 X-ray images with corresponding region annotations. The contribution of our work builds on establishing a multi-modal protocol for our analysis and downstream classification of X-ray images. Our deep learning model trains using data from CT, through generation of synthetic X-ray from available CT scans and including original X-ray images as well. The inference pipleline is exclusively based on X-ray images, not requiring CT data. First we  established a synthetic X-ray generation scheme to generate as many as possible realistic synthetic X-ray to significantly augment X-ray images to expand  our training pool. Second, we use synthetic X-ray as a bridge to transfer the ground-truth from CT to the original X-ray geometry.

\subsection{Method}

\begin{figure*}[!htb]

\centering
\includegraphics[width=0.85\textwidth]{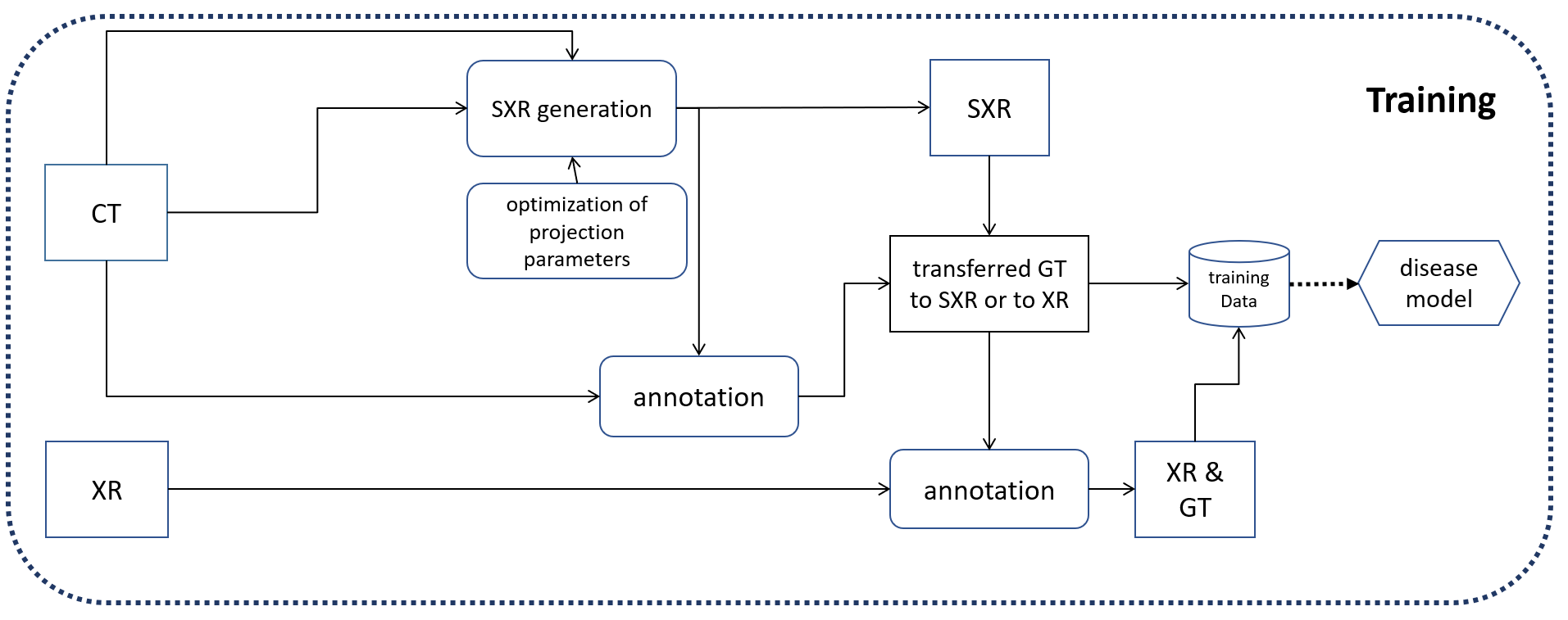}

\caption{The training scheme design \label{fig:training}}
\end{figure*}

\subsubsection{Solution Design Overview}
The key idea is to learn the disease patterns jointly using CT and XR data but to inference the solution on X-ray images. In this study, we are not particularly focusing on network design, but rather on improving data sufficiency and enhancing ground-truth quality to improve the visibility of anomolous tissue on a lower dimensional modality. In order to leverage available unparied and paired X-ray and CT images, we have designed a pipeline as Fig.\ref{fig:training} shows. Disease regions in X-ray and CT are annotated independently by trained staff with varying levels of experience. Using CT, we generate multiple synthetic X-ray image (SXR) per CT volume together with corresponding 2D masks of the diseased region(see \ref{sec:SXR}). For patients where X-rays and CTs (we refer them as paired CT and XR data) are acquired within a small time-window (e.g. 48 hours), we automatically transfer the masks from the SXR to the corresponding XR. The transferred annotations on XR are further adjusted manually reviewed by trained staff to build our pristine mask ground-truth/annotations (see \ref{sec:GTT}). With all available XR and SXR together with corresponding disease masks, we train a deep-learning convolutional neural network for diagnosing and segmenting COVID-19 disease regions on X-ray.

\subsubsection{Synthetic X-ray Generation} \label{sec:SXR}
Original X-ray images are generated  by shining an X-ray beam with $I_0$ initial intensity on the subject and measuring the intensity ($I$) of the beam having passed through an attenuating medium  at different positions using a 2D X-ray detector array. The attenuation of X-ray beams in matter follows the BeerLambert law, stating that the decrease in the beam’s intensity is proportional to the intensity itself ($I$) and the linear attenuation coefficient value($\mu$) of the material being traversed. (\ref{eq:beer_lambert}). 
\begin{equation} \label{eq:beer_lambert}
\frac{dI}{dx} =  -\mu(x)I(x)
\end{equation}
From this we can derive the beam intensity function function over distance to be an exponential decay as the following:
\begin{equation} \label{eq:attenuation}
I=I_{0}e^{-\int\mu(x)dx}
\end{equation}
In medical X-ray images the values saved to file are proportional to the expression seen in \ref{eq:normalized_xray}, which is actually a summation of the attenuation values along the beam. If the subject is made up of smaller homogeneous cells, this could be formulated as a sum.
\begin{equation} \label{eq:normalized_xray}
ln\left(\frac{I_{0}}{I}\right)=\int\mu(x)dx = \sum_{i}\mu_i(x)\Delta x_i
\end{equation}

3D CT images are constructed from many 2D X-ray images of the subject taken at different angles and positions (helical or other movement of the source-detector pair) by using a tomographic reconstruction algorithms such as iterative reconstruction or the inverse radon transform. The voxel values of the constructed CT image are the linear attenuation coefficient values specific to the used X-ray beam’s energy spectrum. These values are converted to Hounsfield unit (HU) values as seen in Eqn \ref{eq:HU_conversion}, where $\mu_{water}$ and $\mu_{air}$ are the linear attenuation coefficient for water and air respectively for the given X-ray source. Using this unit of measure transforms images taken with different energy X-ray beams to have similar intensity values for the same tissues, scaling water to 0 HU, air to -1000 HU.

\begin{equation} \label{eq:HU_conversion}
HU=\frac{\mu-\mu_{water}}{\mu_{water}-\mu_{air}}
\end{equation}

\begin{figure}[!htb]

\centering
\includegraphics[width=0.45\textwidth]{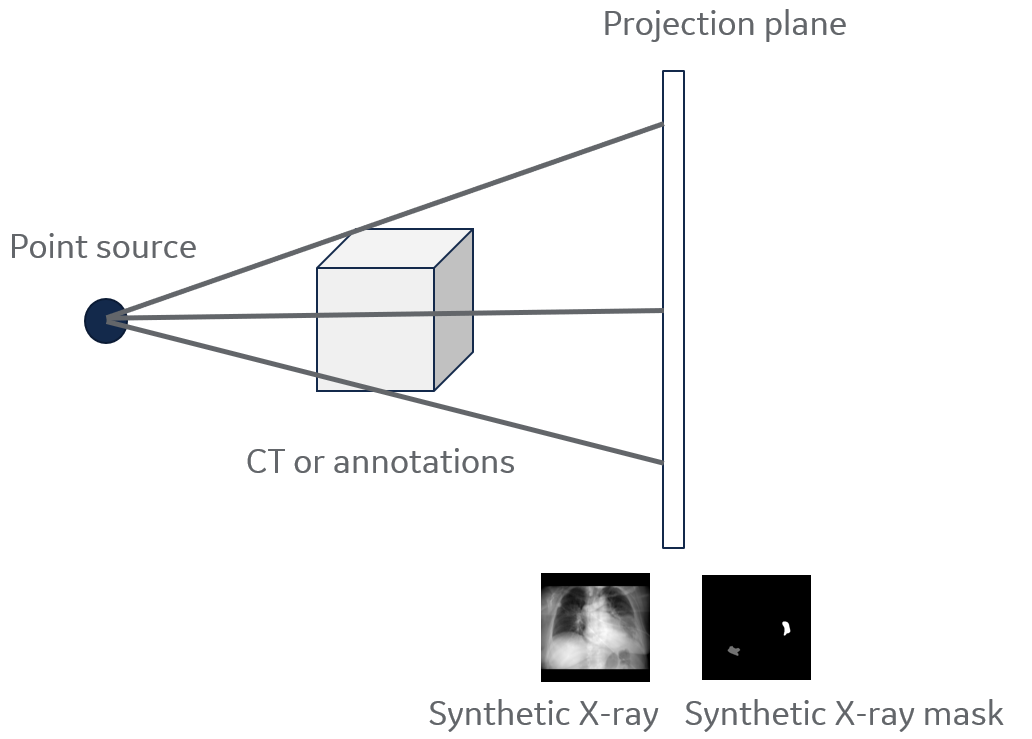}
\caption{The illustration of synthetic X-ray and its mask generation \label{fig:projection}}

\end{figure}

An inverse process can help create X-ray images from CT images by projecting the constructed 3D volume back into a 2D plane along virtual rays originating from a virtual source. Since the X-ray image (Eqn. \ref{eq:normalized_xray}) is an integral/summation of attenuation values along the projection rays, established methods create projection images by converting the CT voxel values from HU to linear attenuation coefficient and simply summing the pixel values along the virtual rays (weighted by the traveled path-length by the ray withing each pixel). To achieve realistic pseudo X-ray we used a point source for the projection which accurately models the source of the X-ray, like in actual X-ray imaging apparatus. For the projection itself we used the ASTRA-toolbox\cite{ASTRA} , which allowed parameterization of the beam’s angle (vertical and horizontal) as well as the distances between the virtual point source, the origin of the subject and the virtual detector (see Fig. \ref{fig:projection}). As X-ray images have the same pixel spacing in both dimensions, while CT images usually have larger pixel spacing in the z direction, we re-sampled the 3D volumes to have isotropic (0.4 mm) spacing prior to computing the projection. In clinical X-ray images, the radiation source is behind and the detector is in front of the patient - also known as posterior-anterior (PA) view the images generated appear to be flipped horizontally. To conform with this protocol, we also flipped PA projections.

It is important to point out that in case of small field-of-view CT images, where the patient doesn't fit into the reconstruction circle in the horizontal plane, creating coronal or sagittal projection will not result in lifelike X-rays image as body parts are missing from the virtual ray's path.

During our work we also projected the segmentation ground truth masks for CT images to 2D format. The same transformations were applied to these binary masks as to the corresponding CT volumes excluding the ones aiming to scale the voxel/pixel values.
In our workflow we use both a binary format of the 2D projected masks - created by setting all non-zero values to 1 - and a depth-mask format which retains information of the depth of the original 3D mask. The latter was scaled so the values have the physical meaning of depth.

\subsubsection{Pristine Annotation Generation}  \label{sec:GTT}
For paired XR and CT from same patients, our aim is to transfer the pixel-wise ground-truth from CT to SXR and from SXR to XR. SXR serves as a bridge between CT and XR.

For each CT volume, we generate a number of SXRs and corresponding disease masks by varying imaging parameters as mentioned in \ref{sec:SXR}. The goal is to register each SXR candidate to XR so that the disease region between SXR and XR have the best alignment so that we can apply the same transformation on the disease mask from SXR and generated registered disease mask for the paired XR. As for the same CT, multiple SXRs are generated resulting different registered masks and we simply select the one with the best mutual information between registered SXR and XR. Another approach to select optimal SXR involves comparison of the lung mask region. The pair with the most overlapping mask was used to select the ideal pair.

\begin{figure*}[!htb] 
    \begin{subfigure}[b]{0.35\linewidth}
        \includegraphics[width=50mm]{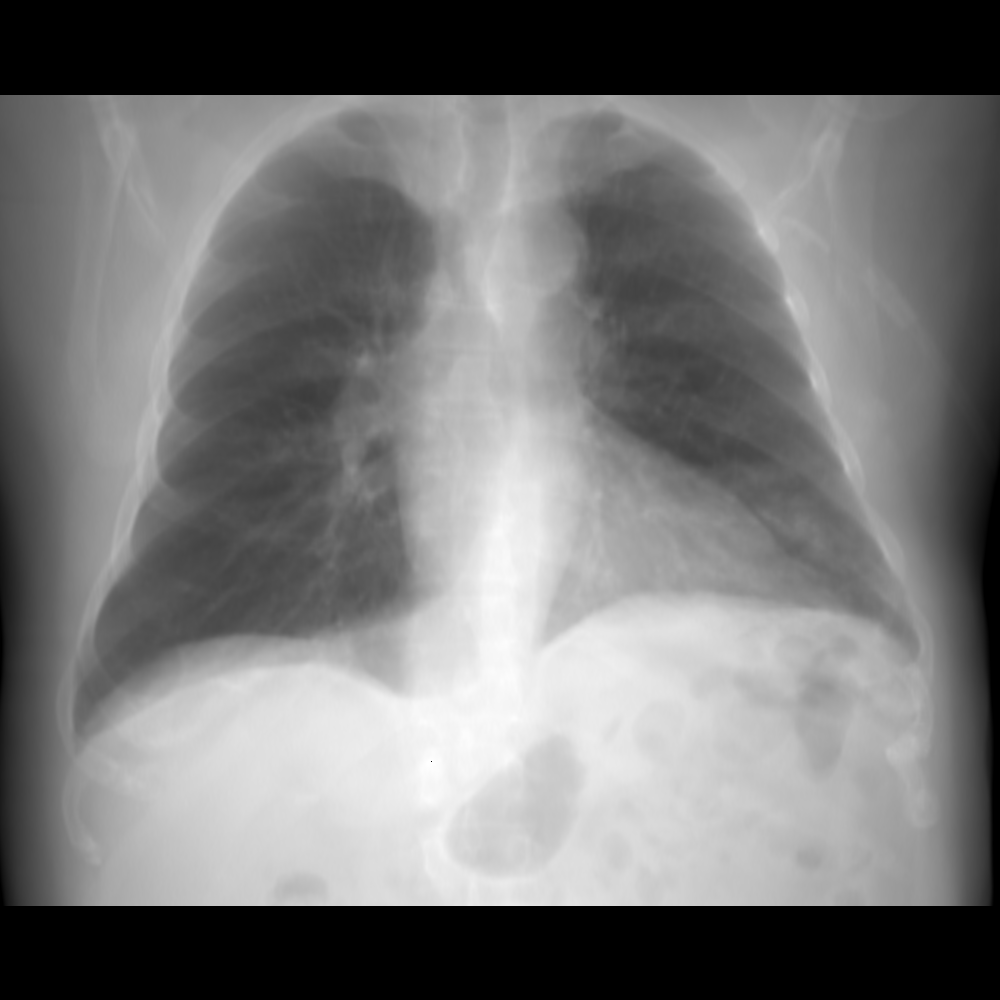}
        \caption{}
    \end{subfigure} %
    \begin{subfigure}[b]{0.35\linewidth}    
        \includegraphics[width=50mm]{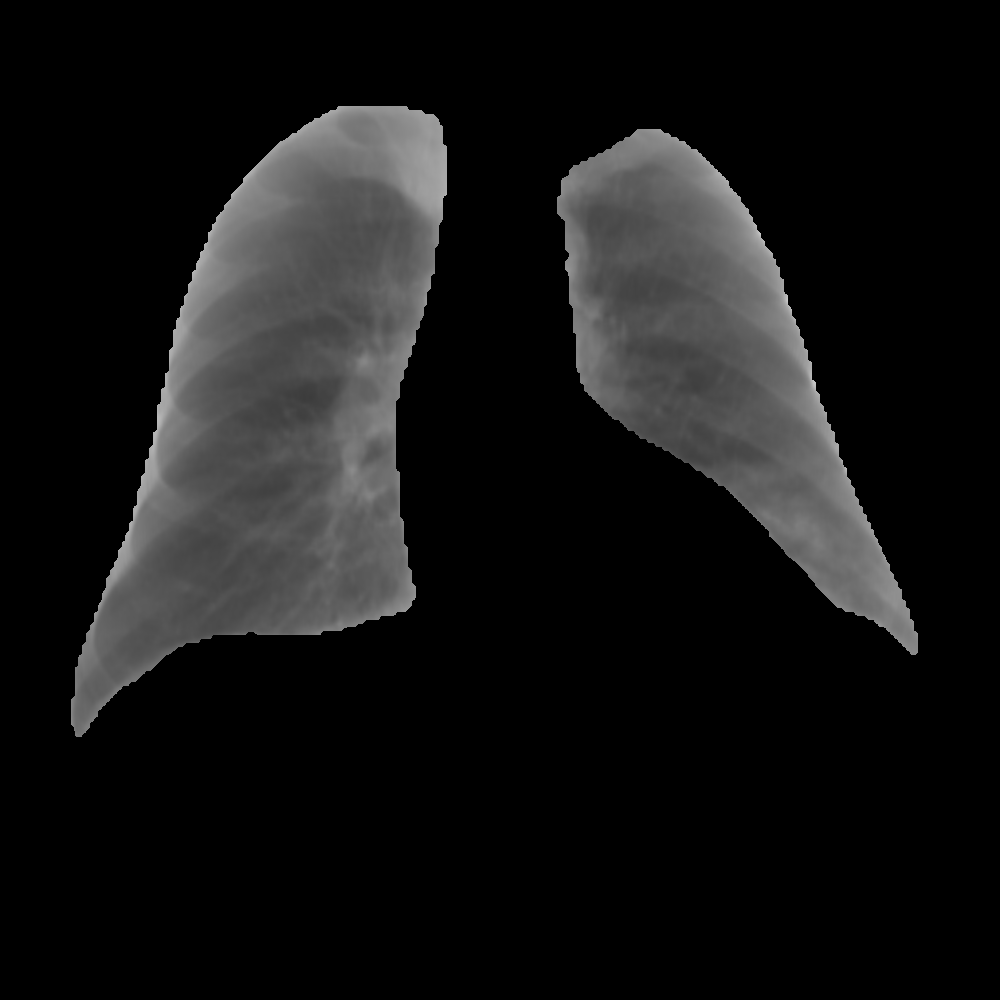}
        \caption{}
    \end{subfigure} 
    \begin{subfigure}[b]{0.35\linewidth}    
        \includegraphics[width=50mm]{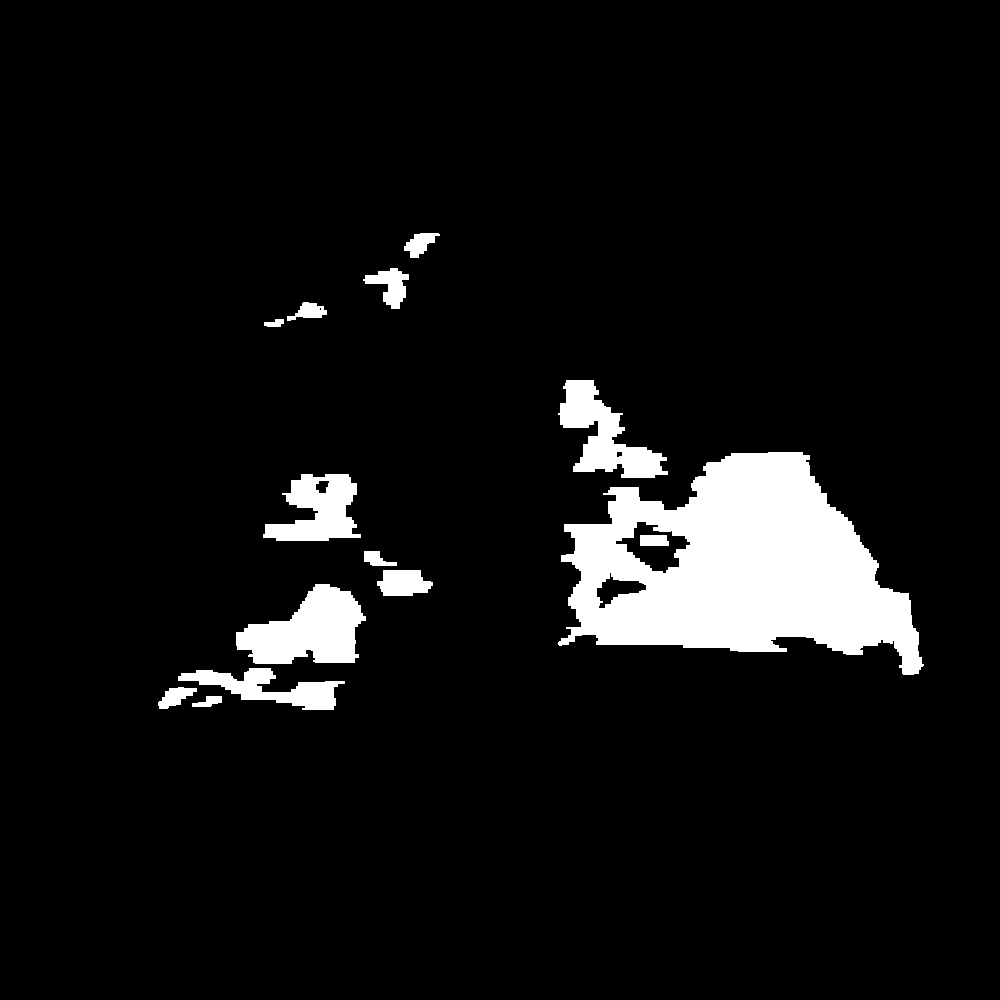}
        \caption{}
    \end{subfigure}  

     \begin{subfigure}[b]{0.35\linewidth}
        \includegraphics[width=50mm]{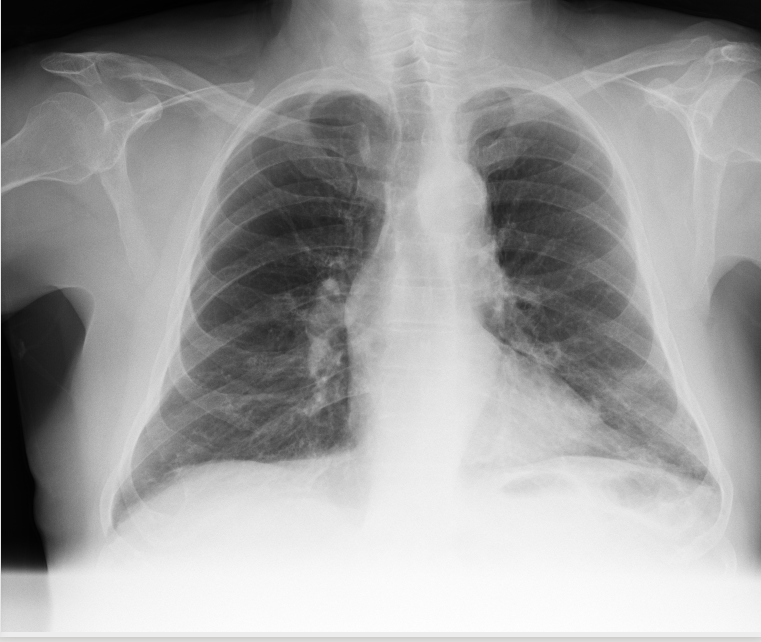}
        \caption{}
    \end{subfigure} %
    \begin{subfigure}[b]{0.35\linewidth}    
        \includegraphics[width=50mm]{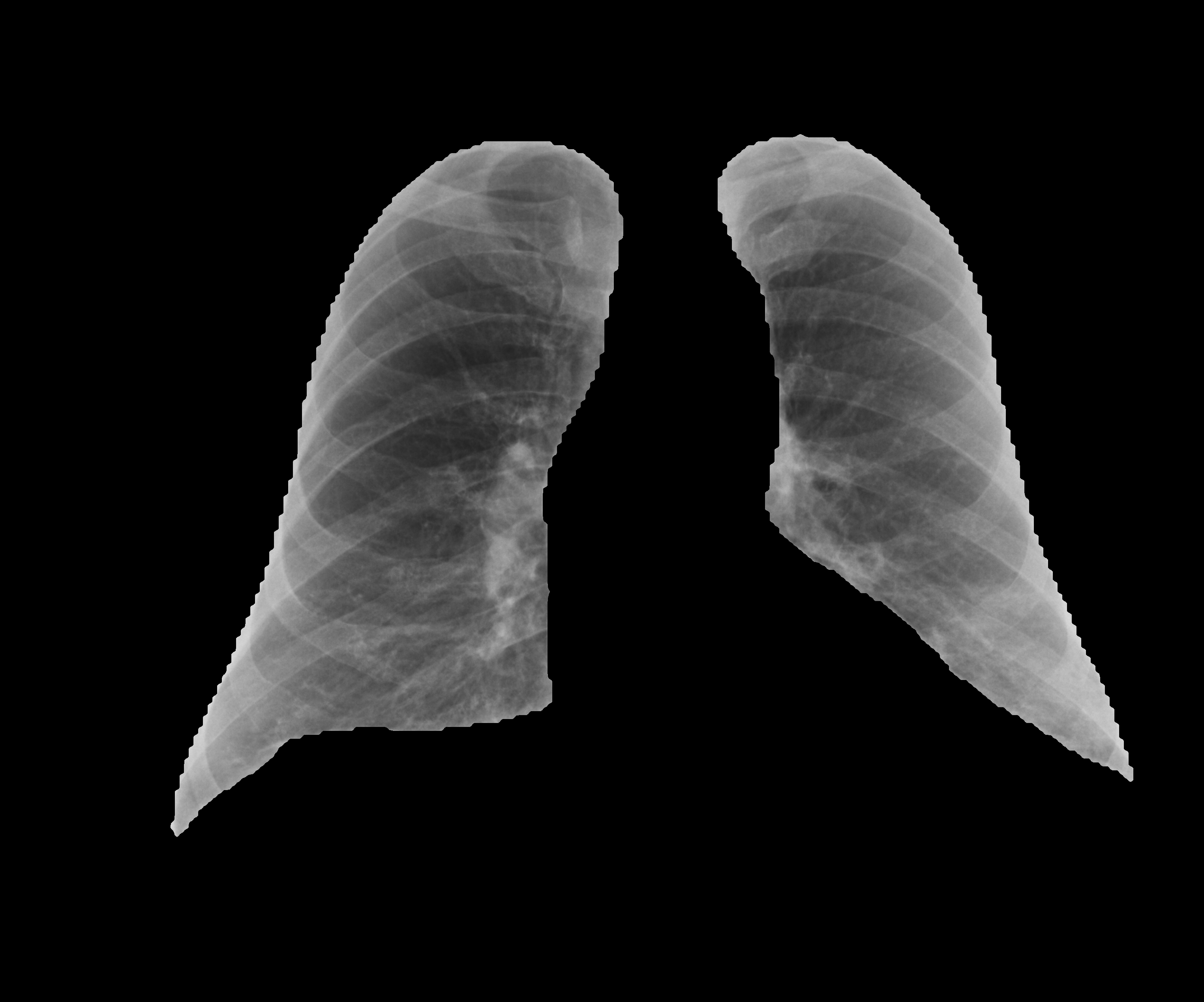}
        \caption{}
    \end{subfigure} 
    \begin{subfigure}[b]{0.35\linewidth}    
        \includegraphics[height=42mm, width=50mm]{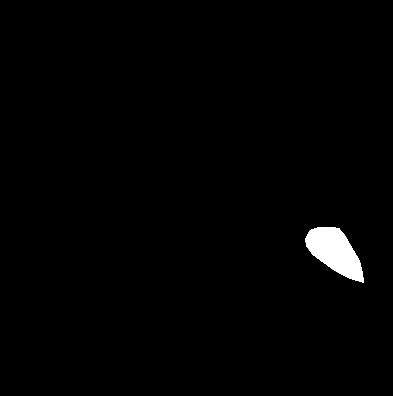}
        \caption{}
    \end{subfigure}   
    
    \begin{subfigure}[b]{0.35\linewidth}
        \includegraphics[width=50mm]{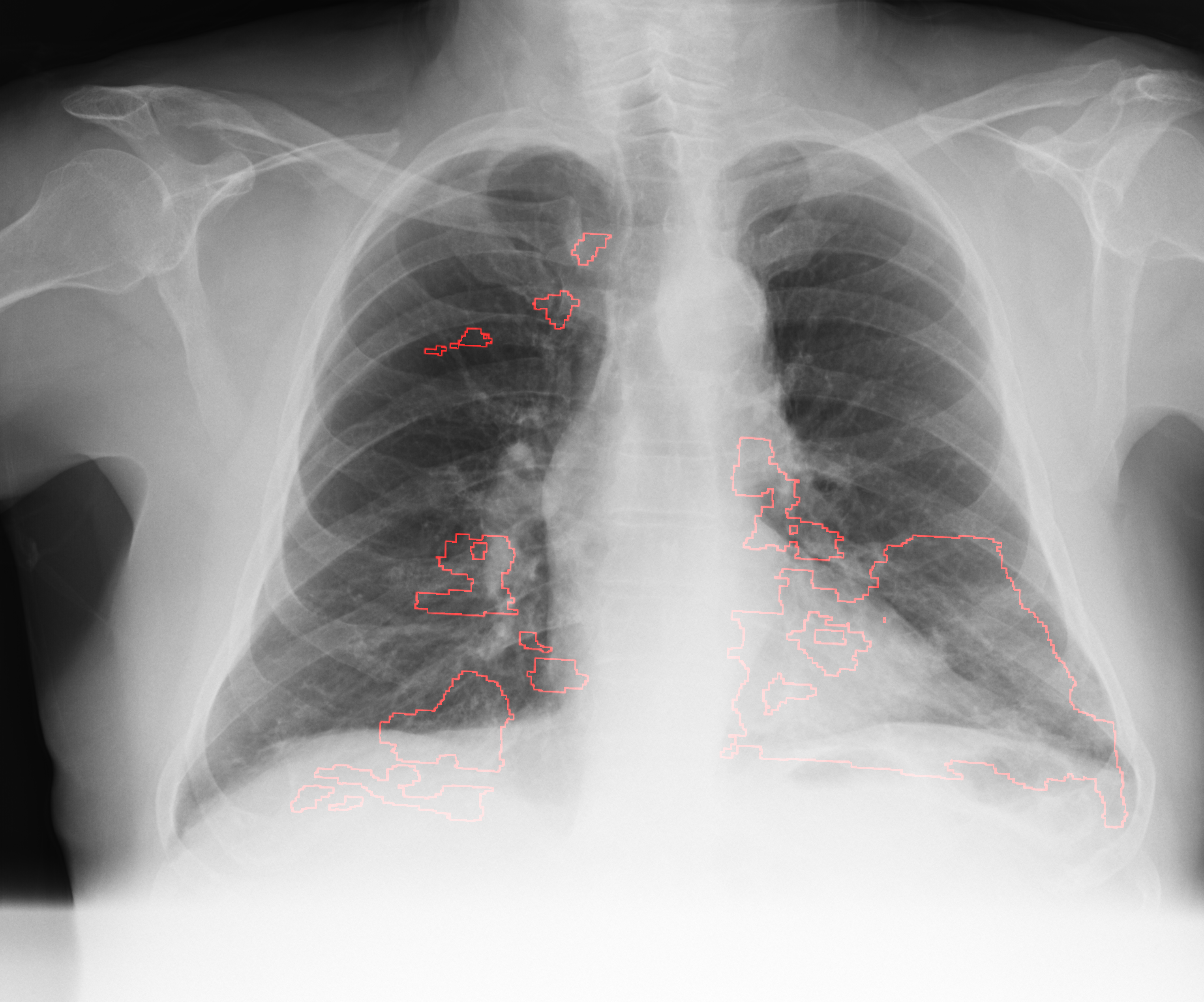}
        \caption{}
    \end{subfigure} %
    \begin{subfigure}[b]{0.35\linewidth}    
        \includegraphics[width=50mm]{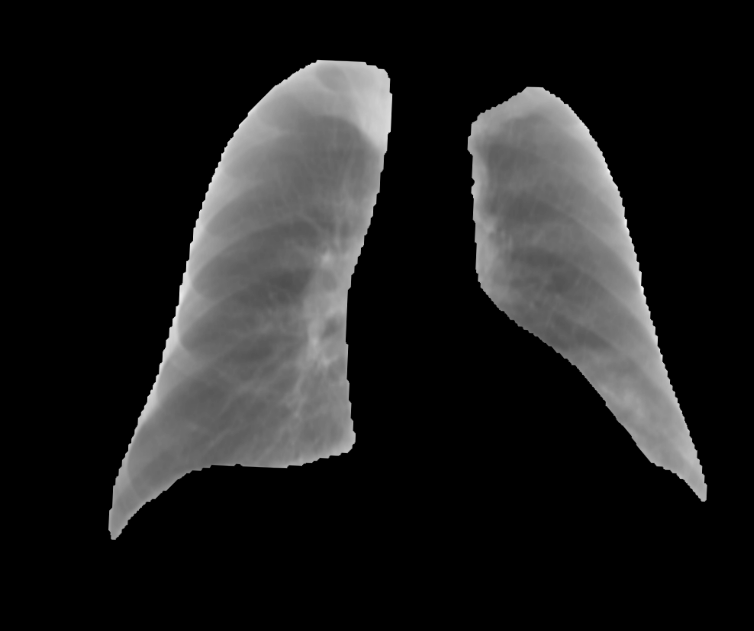}
        \caption{}
    \end{subfigure} 
    \begin{subfigure}[b]{0.35\linewidth}    
        \includegraphics[width=50mm]{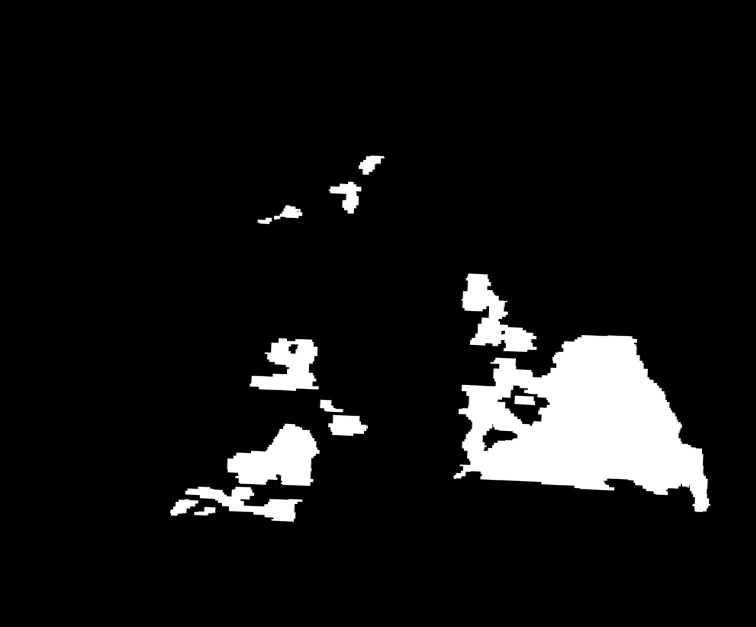}
        \caption{}
    \end{subfigure}

    \caption{An example of transfering synthetic X-ray mask to X-ray mask. Top: a representative synthetic X-ray generated from CT, the corresponding lung image and disease mask; Middle: paired X-ray, the corresponding lung image and direct disease annotation from X-ray; bottom: X-ray with transferred annotations from CT shown as red contour; registered lung image from synthetic X-ray and transferred disease annotations from synthetic X-ray}
    
\label{fig:reg}    
    
\end{figure*}

To register SXR to XR, one challenge is that the fields of body view are different between the two and SXR is imaged usually with hands and arms up while for XR the hands are down, normal position of the body. To alleviate the problem of the during image registration, instead of using original images, we create lung ROI image by applying lung segmentation on synthetic X-ray and X-ray. Fig. \ref{fig:reg} shows one example how we transfer the CT ground-truth from CT to synthetic X-ray and from synthetic X-ray to real X-ray. We observe that there are moderate differences between annotations from original X-ray  and those annotations derived from transformed and transferred CT data implying that the visibility of COVID-19 related pneumonia on X-ray may not be ideal and most comprehensive.

Although we have an approximate spatial match between SXR and XR, we cannot directly use transferred groundtruth(GT) from 3D CT onto 2D even with the best matching for training the deep learning models. The disease can rapidly change even within 48 hours. In this case the transferred groundtruth from CT can only be used as a directional guidance for a human expert to annotate on the 2D X-rays. Therefore for paired cases, we perform manual adjustment on the automated transferred ground-truth.

\begin{figure*}[!htb]

\centering
\includegraphics[width=0.85\textwidth]{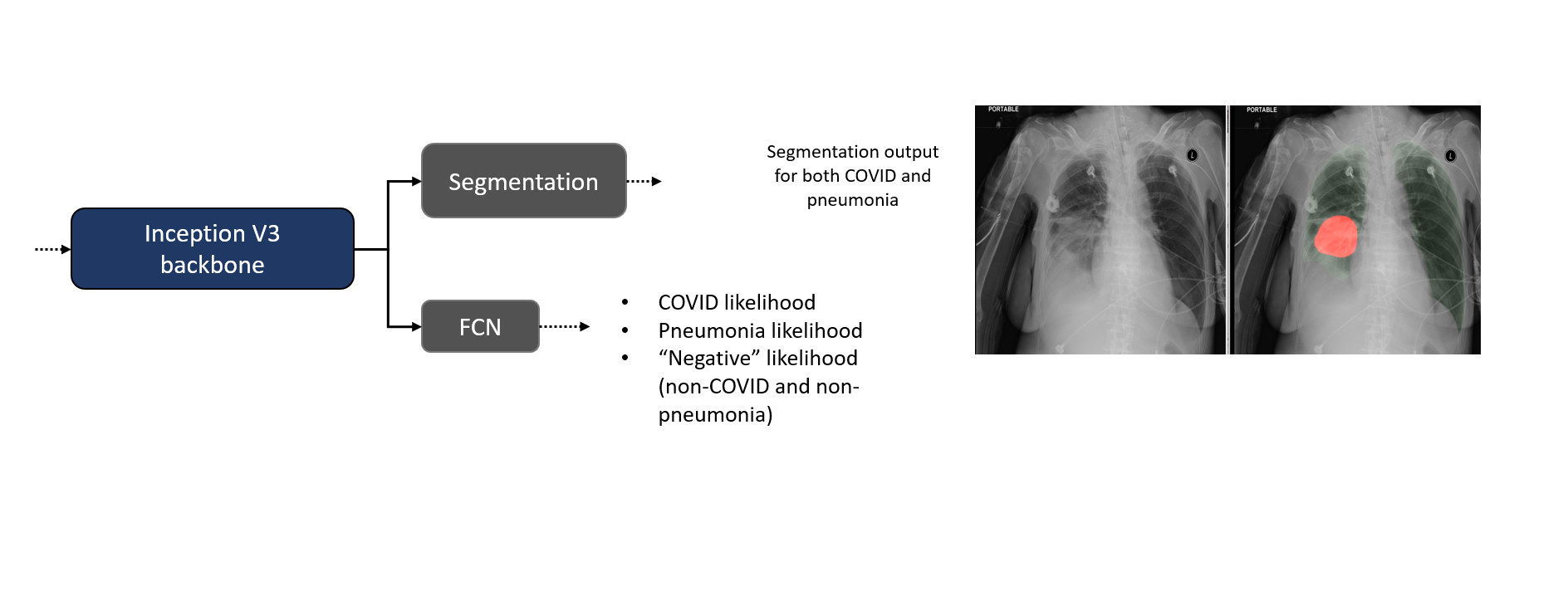}
\caption{The schematic overview of our proposed classification and segmentation model \label{fig:covid_model}}

\end{figure*}

\subsubsection{COVID-19 Modeling and Evaluation Strategy}

For this extra-ordinary COVID-19 pandemic, AI systems have been investigated to identify anomalies in the lungs and assist in the detection, triage, quantification and stratification (e.g. mild, moderate and severe) of COVID-19 stages. To help radiologists do the triage, our deep-learning model takes frontal (anterial-posterior view or posterior-antierial view) X-ray as input and output two types of information (see Fig. \ref{fig:covid_model}): location of the disease regions and classification. For location, the network generates a segmentation mask to identify disease pixels or regions on X-ray related to COVID-19 or regular pneumonia. For classification, the fully connected neural network (FCN) outputs the likelihood of COVID-19 pneumonia, other pneumonia or a negative finding for each given input AP/PA X-ray image.

Many publications have released their algorithms in open online forums or are marketing the same as additional pneumonia indicators. However, the robustness of the algorithms and their clinical value is somewhat unproven. A few studies have characterized systems for COVID-19 prediction with stand-alone performance that approaches that of human experts. However, all the existing works have either no established pixelwise ground-truth or are evaluated using pixel-wise ground-truth from purely X-ray annotations with uncertainties from annotators

Our deep learning model was trained with and without using data from CT cases. We evaluated our approach in three seperate aspects: first, AI predictions were compared with the image-level classification labels; second, the segmentation of disease regions for the COVID class was evaluated against direct X-ray pixel-wise annotations. Third, the segmentation of disease regions for the COVID-19 class was evaluated against pristine pixel-wise annotations.

\subsection{Data and Groundtruth}

In this study, we formed a large experiment dataset consisting of real X-ray images and synthetic X-ray images originating from CT volumes. The data was sourced  from in-house/internal collections as well as publicly available data sources including Kaggle Pneumonia RSNA \cite{KagglePneumonia}, Kaggle Chest Dataset \cite{KaggleChest}, PadChest Dataset \cite{bustos2019padchest}, IEEE github dataset\cite{IEEE-Github}, NIH dataset \cite{wang2017chestx}. Any image databases with limiting non-commercial use licenses were excluded from our train/test cohorts. Representative paired and unpaired CT and XR datasets from US, Africa, and European population were also leveraged and sourced through our data partnerships. Outcomes were derived from information aggregated from radiological and laboratory reports. A summary of the database and selected categories used for our experiments is summarized in Table \ref{table:alldata} where the in-house test dataset is a subset of test dataset. As the trained model will be inferenced on real X-ray images, we remove synthetic X-rays for validation and testing.

\begin{table*}
\begin{center}
    \begin{tabular}{| l |  l | l |}
    \hline
    Dataset & X-rays (\# XMA , \# PMA) & Synthetic X-rays (\# SMA)\\ \hline
    train COVID-19 & 974 (247, 77) & 21487 (8322) \\ 
    train pneumonia & 10175(6108, 17) & 11312 (5380) \\ 
    train negative & 17068 (NA,NA) & 12542 (NA) \\ \hline
    val COVID-19 & 113 (37,2) & NA \\ 
    val pneumonia & 531 (473,8) & NA \\ 
    val negative & 3731 (NA, NA) & NA \\ \hline    
    test COVID-19 & 307 (68, 52) & NA \\ 
    test pneumonia & 1006 (345,33) & NA \\ 
    test negative & 2271 (NA,NA) & NA \\ \hline
    in-house test COVID-19  & 268 (68,52) & NA \\ 
    in-house test pneumonia   & 119 (45,33) & NA \\ 
    in-house test negative & 34 (NA,NA) & NA \\ \hline
    \end{tabular}
    
    \caption{Dataset breakdown for our experiments \label{table:alldata}}
\end{center}

\end{table*}

Two levels of groundtruth are associated with each image: image-level groundtruth and pixel-level (segmentation) groundtruth. 

For image-level groundtruth, each image is assigned with a label of COVID-19, pneumonia or negative. All in-house X-rays and CTs and the COVID-19 images from the public data sources were confirmed by PCR tests. The labels of pneumonia and normal images from public data sources are given by radiologists.

For pixel-wise/segmentation groundtruth, we have four types of annotations. X-ray manual mask annotations (XMA) are made by annotators purely based on X-rays without any information from CTs.  Synthetic mask annotations (SMA) are generated by the projection algorithms based on CT annotations for synthetic X-rays. Transferred CT mask annotations (TMA) are automatically generated by registration algorithms which transfer annotations from CT to X-ray using SXR as a bridge. Pristine mask annotations (PMA) which are  adjusted annotations by annotators from TMA.

\subsection{Experiment Settings}
To show the benefits of multi-modal training for developing COVID-19 model, we have conducted training with 4 different training datasets summarized in Table \ref{table:strategies}.

\begin{table*}
\begin{center}
    \begin{tabular}{| l | l | }
    \hline
    training set & Description \\ \hline
    T1 & X-ray images with XMA \\ \hline
    T2 & T1 + synthetic X-ray images with SMA  \\ \hline
    T3 & T1 + X-ray images with PMA \\ \hline
    T4 & T1 + synthetic X-ray images with SMA + X-ray images with PMA\\ \hline
    \end{tabular}
    \caption{Different training sets \label{table:strategies}}
\end{center}
\end{table*}

\subsection{Evaluation Metrics}
To evaluate the performance of our inferred classification of the test subjects, we use the area under the receiver operating characteristic curve (AUC) between different combinations of  positive  negative classes including COVID-19 vs pneumonia, COVID-19 vs pneumonia+negative and COVID19 vs negative for different usage scenarios.

$Dice$ coefficient is used to evaluate the exactness of our pathology localization,  

\subsection{Results}

\subsection{Pristine annotation creation}
With three different pixel-wise annotations on X-ray, we evaluated overlap between XMA, PMA and TMA.

\begin{table}
\begin{center}
    \begin{tabular}{| l | l | }
    \hline
    Comparisons & $Dice$ \\ \hline
    XMA vs TMA & 0.28  \\ 
    PMA vs TMA & 0.47 \\ 
    XMA vs PMA  & 0.50  \\ \hline
    \end{tabular}
    \caption{Area overlapping between different annotations \label{table:ground}}
\end{center}
\end{table}

\begin{figure*}[!htb]
\centering
\includegraphics[width=0.85\textwidth]{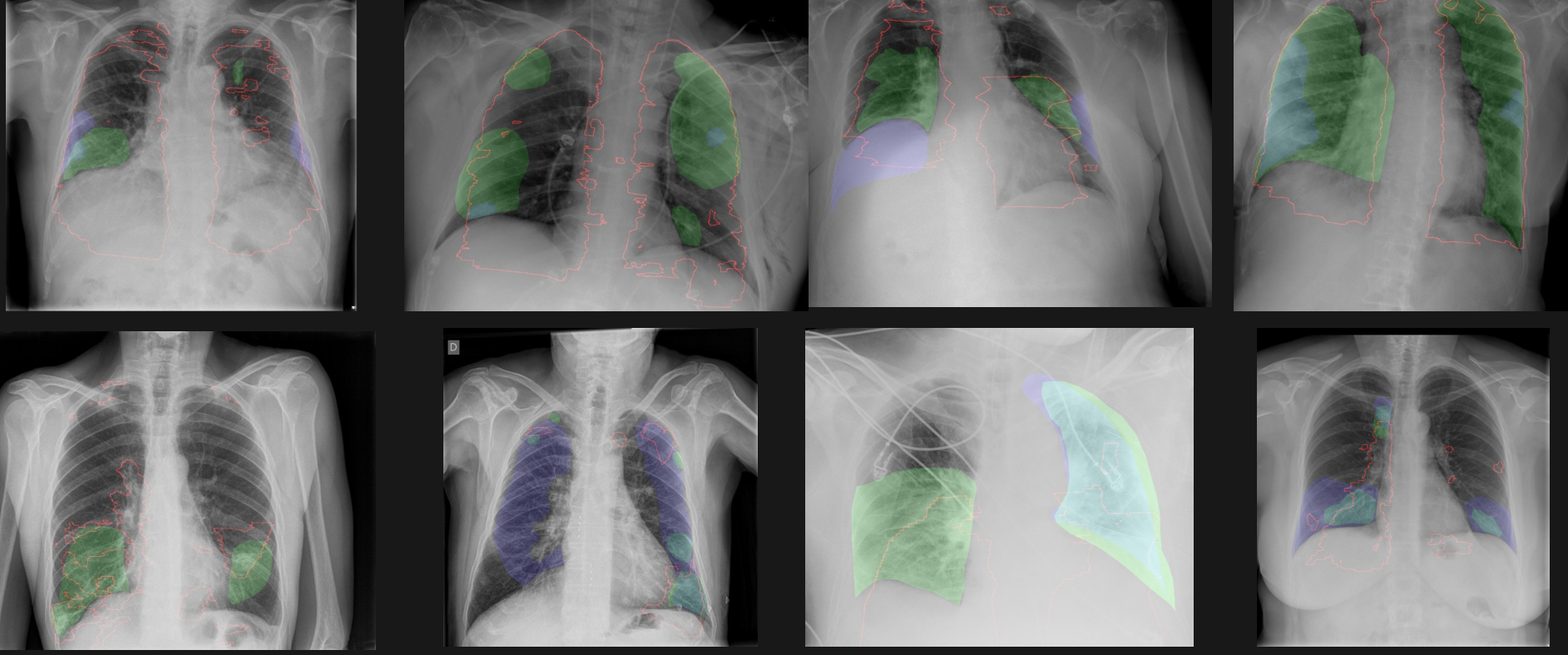}
\caption{Examples with TMA as red contour, XMA as blue regions and PMA as green regions \label{fig:annotationTransfer}}
\end{figure*}

We can observe (see Table \ref{table:ground}) that after using TMA (transferred CT annotations), the consistency of human annotations to CT annotations is largely improved from 0.28 to 0.47 in terms of $Dice$ coefficient. The $Dice$ coefficient between XMA (X-ray manual mask annotations) and PMA (pristine mask annotations) are also moderate which means PMA is somewhere between X-ray direct annotations and CT transferred annotations. Fig.\ref{fig:annotationTransfer} shows a number of examples with TMA, XMA and PMA. The X-ray annotations show large inconsistency with automated CT transferred annotations.

\subsection{Model Evaluation}

The evaluation of the model was performed in terms of both classification and segmentation of COVID-19 disease regions. To test the effect of domain shift, we show the evaluation results on both all X-ray test dataset and in-house test dataset where COVID19 regular pneumonia and normal cases are all available.

Regarding classification, Table \ref{table:classification} shows AUC based on different combination for positive and negative classes. By adding synthetic X-rays, the AUC increases for COVID-19 vs pneumonia + negative increase from 0.89 to 0.93. The addition of adding pristine groundtruth does not further increase the AUC.  The same increase is observed if AUC is computed using COVID-19 as positive and pneumonia as negative class.

\begin{figure*}[!htb]

\centering
\includegraphics[width=0.85\textwidth]{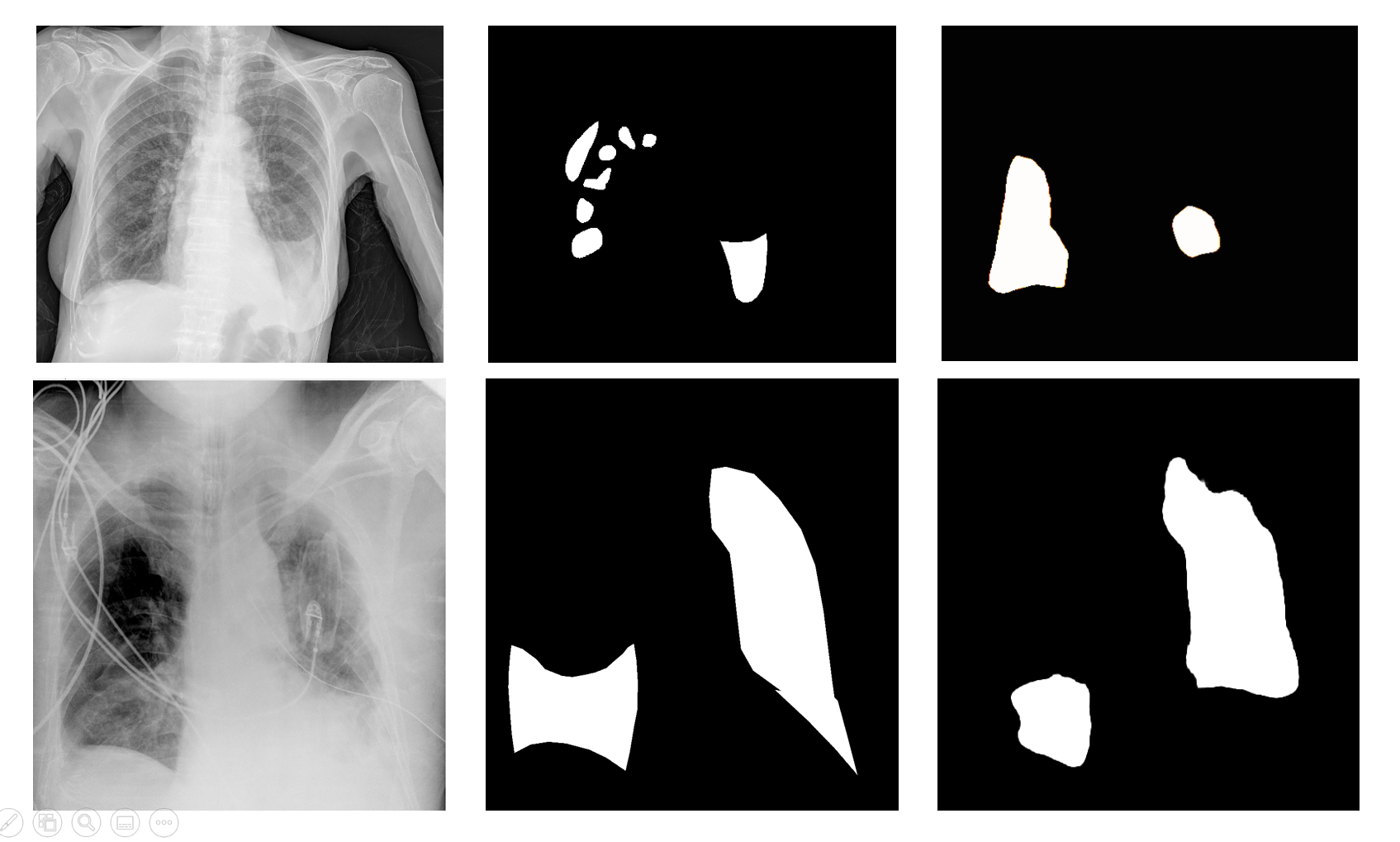}

\caption{Segmentation examples where images on the left are original X-ray images, in the middle are PMA and on the right are AI segmentations \label{fig:segmentation_ex}}

\end{figure*}

We measure the Dice coefficients to estimate the segmentation accuracy. As PMA are obtained, we have measured Dice on all testing images, in-house test images and testing images with PMA only shown by Table \ref{table:Dice}. Adding PMA can largely improve the $Dice$ measures across different test settings up to 0.70. Fig.\ref{fig:segmentation_ex} shows examples of AI detection and segmentation of COVID-19 regions with manual annotations as overlay as well.

\begin{table*}
\begin{center}
    \begin{tabular}{| l | l | l | l| }
    \hline
    Train dataset  vs  test Dice  & XR with XMA  & in-house XR with XMA  & in-house XR with PMA\\\hline
    T1  & 0.58 & 0.59 & 0.59 \\ \hline
    T2  & 0.57 & 0.56 & 0.58 \\ \hline
    T3 & \textbf{0.60} & \textbf{0.62} & \textbf{0.70} \\ \hline
    T4 & 0.57 & 0.58 & 0.62 \\ \hline
    \end{tabular}
    
    \caption{Segmentation results: dice measures of different training schemes on different datasets \label{table:Dice}}
\end{center}
\end{table*}

\begin{table*}
\begin{center}
    \begin{tabular}{| l | l | l | l| l|}
    \hline
    Training  vs  test AUC  & \makecell{ AUC C vs P+N \\ on XR}  &  \makecell{ AUC C vs P \\ on XR}  &  \makecell{ AUC C vs P+N \\ on in-house XR} & \makecell{AUC C vs P \\on in-house XR} \\\hline
    T1  & 0.98 & 0.98 & 0.89 & 0.87\\ \hline
    T2  & \textbf{0.99} & 0.98 & \textbf{0.93} &  \textbf{0.92}\\ \hline
    T3 & \textbf{0.99} & 0.98 & 0.91 & 0.90 \\ \hline
    T4 & \textbf{0.99} & \textbf{0.99} & \textbf{0.93} & \textbf{0.92}\\ \hline
    
    \end{tabular}
    
    \caption{Classification results: AUC measures of different training schemes on different datasets with different positive and negative compositions\label{table:classification}}
\end{center}
\end{table*}

\subsection{Conclusion and Discussion}

In this study, from multi-modal perspective, we have developed an artificial intelligence system which learns from high dimensional modality CT but inferences on low dimensional mono-modality X-ray for COVID-19 diagnosis and segmentation. The system classifies a given image into three categories: COVID-19, pneumonia and negative. We show that by learning from CT, the performance of the AI system is improved in terms of both classification and segmentation. Our AI system achieves a classification AUC of 0.99 and 0.93 between COVID-19 and pneumonia plus negative on full testing dataset and the subset in-house dataset, respectively. The Dice of 0.57 and 0.58 are obtained for COVID-19 disease regions on full testing dataset and the subset in-house dataset using X-ray direct annotations, respectively. The Dice is increased to 0.62 when pristine ground-truth transferred from CT is used for training and testing. Accurate classification can aid physician in triaging patients and make appropriate clinical decisions. Accurate segmentation enhanced confidence in the triage, and helps in quantitative reporting of disease for reporting and monitoring progression.

Learning from a second modality (CT) in our multi-modal approach has two main implications. One impact is to add synthetic X-ray and corresponding disease masks with different projection parameters to significantly augment training image pool to ensure data diversity. Another benefit related to additional pathology evidence from a higher sensitivity modality when we transfer the CT annotations from  synthetic X-ray, and from synthetic X-ray to original X-ray, and allowing manual fine tuning and  the annotations used for training. The first addition contributes mostly towards gain in the classification accuracy and the second addition contributes substantially to the gain in disease localization. The quality of the synthetic X-ray may play an important role and is worth further investigation, perhaps using generative networks to make the synthetic X-ray more realistic.

Our study shows that there exists substantial inconsistency between X-ray direct annotations and automated transferred CT annotations. When using transferred annotations as hints, the second version of X-ray annotations (pristine groundtruth) are more consistent to automated transferred CT annotations. The Dice between direct annotations and pristine groundtruth is 0.50 which might be a good reference indicating how good segmentation the AI should achieve. The pristine groundtruth is something between X-ray annotations and CT annotations. It should be noted that the automated transferred annotations can not be directly used as errors might be induced because of registration. Also COVID-19 is a fast changing disease, as we are pairing X-ray and CT acquired within a time-window of 48 hours, the disease status in CT might be quite different compared to the disease status when X-ray is taken. Therefore manual adjustment is an important consideration.

In the research and industry community, efforts have been made to apply AI into imaging-based pipeline of for the COVID-19 applications. However, many existing AI studies for segmentation and diagnosis are based on small samples and based on single data source, which may lead to the overfitting of results. To make the results clinically applicable, a large amount of data from different sources shall be collected for evaluation. Moreover, many studies only provide classification prediction without providing segmentation or heatmap which makes AI systems lack  explainability. By providing the segmentation, we aim to fill this void, enhancing the promotion of AI in clinical practice. On the other hand, the imaging-based diagnosis has limitations and clinicians make the diagnosis considering clinical symptoms also. An AI system can be largely enhanced with incorporation of patient clinical parameters \cite{mei2020artificial} such as blood oxygen level, body temperature, to further enhanced the capability to accurately diagnose pathological conditions.

WHO has recommended\cite{who2020-chest} a few scenarios where chest-imaging can play an important role in care delivery. From the triage perspective, WHO suggests using chest imaging for the diagnostic workup of COVID-19 when PCR testing is not available (timely) or is negative while patients have relevant symptoms. In this case, the classification support from our AI can aid radiologists to identify COVID-19 patients. From monitoring or temporal perspective, for patients with suspected or confirmed COVID-19, WHO suggests using chest imaging in addition to clinical and laboratory assessment to decide on hospital admission versus home discharge, to decide on regular admission versus intensive care unit (ICU) admissions, to inform the therapeutic management. In these scenarios, accurate segmentation of disease regions is essential for the evaluations.

The COVID-19 disease continues to spread around the whole world. Medical imaging and corresponding artificial intelligence applications together with clinical indicators provides solutions for triage, risk analysis and temporal analysis. This study provides a solution from multi-modal perspective to leverage the CT information but to inference on X-ray to avoid the necessity of taking CT imaging because of limited accessibility, dose and decontamination concern. Future work focusses on leveraging paired CT information to estimate severity and other higher dimensional measures. We believe that our study introduces a new trend of combinating  multi-modal training and single-modality inference. Although this study tries to leverage CT information as much as possible to aid the data-driven AI solution on X-rays, it should be noted that the extraction of relevant information is limited by the nature of X-ray imaging because of 2D projection as well as impaired visibility with the presence of the non-lung thick tissue. In addition different vendors may apply varying post-processing algorithms to suppress information on those thick tissue regions. To avoid information loss, our AI may be deployed directly on X-ray hardware with direct access to X-ray raw images for ideal translation to the clinic.

\bibliographystyle{unsrtnat}
\bibliography{egbib}

\end{document}